\documentclass[sn-apa]{sn-jnl}

\usepackage{xcolor}   
\usepackage{manyfoot} 

\usepackage{graphicx}
\usepackage{booktabs}
\usepackage{makecell}
\usepackage{multirow}
\usepackage{subcaption}
\usepackage{amsmath}

\begin{document}

\title[Digital Lifelong Learning in the Age of AI]{Demographic Trends in Digital Lifelong Learning in the Age of AI}

\author*[1]{\fnm{Geeta} \sur{Puri}}\email{geeta\_puri@mymail.sutd.edu.sg}
\author[1]{\fnm{Nachamma} \sur{Sockalingam}}
\author[1]{\fnm{Dorien} \sur{Herremans}}

\affil[1]{\orgname{Singapore University of Technology and Design}, \orgaddress{\street{8 Somapah Road}, \postcode{487372}, \country{Singapore}}}

\abstract{Digital platforms and AI tools are used for learning across educational and everyday settings, but learners may value different features. We analysed two anonymous online surveys recruited through social networks: Survey~A contained 118 usable responses collected in 2024, and Survey~B contained 81 responses collected in 2025. The surveys were analysed separately because respondent overlap could not be determined. Survey~A examined reported platform use, recalled pre- and post-pandemic relevance, and valued learning features; Survey~B examined AI use and perceptions. In Survey~A, 84.7\% reported using digital learning materials more frequently than before COVID-19, and recalled relevance was higher for the post-pandemic period (Wilcoxon signed-rank test, $p<0.001$). Ease of use was selected by 89 of 115 respondents (77.4\%), while gamification-motivation ratings did not differ significantly across age groups ($p=0.81$). In Survey~B, 50 of 79 respondents (63\%) selected inaccurate responses as a concern about AI tools. These exploratory findings describe reported experiences and preferences; convenience sampling, small subgroups, and retrospective self-report limit generalisation and causal interpretation.}

\keywords{Lifelong Learning, AI in Education, Gamification, Workforce Development, Digital Learning Platforms, Adult Learning}

\maketitle

\section{Introduction}\label{sec:intro}

Digital platforms support learning in formal education, workplaces, and everyday settings. The pandemic expanded reliance on remote instruction, while recent AI tools have introduced additional ways to seek explanations, generate content, and obtain feedback \citep{khamis2021covid,yan2024practical}.

Research on digital learning spans schools, universities, and older-adult education. A systematic review of digital-literacy programmes for older adults highlights the need to consider practical barriers, prior digital competence, and geographic diversity \citep{xu2025lifelong}. Our study complements this literature by examining reported platform use and learning-feature preferences across age groups and countries.

Three questions motivate this study: how reported platform use varies across demographic groups, which features learners associate with motivation, and how they perceive AI tools. These questions concern reported experiences and preferences; the study does not evaluate the effectiveness of platforms or gamification interventions.

Age may be associated with digital experience, but it is not a direct measure of digital competence. Research has challenged the assumption that younger people form a uniformly proficient generation of digital learners \citep{bennett2008digital,ng2012digitalnatives}. We therefore use age groups descriptively and do not infer digital skills or career stage from age alone.

We examine reported platform use, valued learning features, and perceptions of AI tools across age groups and countries. The findings are intended to inform hypotheses for future platform design and evaluation. Specifically, this study aims to:
\begin{itemize}
    \item \textbf{Map demographic digital-learning engagement:} describe reported digital-learning use and recalled relevance across age groups, countries, and genders (RQ1).
    \item \textbf{Describe platform use:} identify which digital learning platforms respondents in each age group and country reported using (RQ1).
    \item \textbf{Assess AI's role:} describe reported uses of AI tools and the features respondents value or dislike (RQ2).
    \item \textbf{Examine motivations and engagement preferences:} describe the platform features respondents value and how they perceive gamification in relation to motivation across demographic segments (RQ3).
\end{itemize}

The next section reviews research on digital learning, AI tools, and gamification and states the research questions. Section~\ref{sec:methodology} details the methodology, participant characteristics, and survey design. Section~\ref{sec:results} presents reported platform use, valued learning features, and perceptions of AI and gamification across demographic groups. Section~\ref{sec:discussion} discusses the findings for each research question, and Section~\ref{sec:conclusion} concludes with limitations and directions for future research.

\section{Related Literature}\label{sec:lit_review}

The pandemic accelerated the use of remote teaching and digital resources \citep{khamis2021covid}. Research now examines a range of educational technologies, including AI-based support and cyber-physical learning \citep{yan2024practical,Sockalingam2025CyberPhysical}. Their educational value depends on implementation and context, making it important to distinguish technological capabilities from demonstrated learning outcomes.

\subsection{Online Learning Across Demographics}

In a school-wide comparative study at one US dental school, \citet{zheng2021online} found that course performance during online instruction in summer 2020 was comparable to or higher than performance in the corresponding face-to-face courses in summer 2019. However, digital inequality remains a persistent barrier. A systematic review of digital-literacy training for older adults identified disparities in access to learning resources and highlighted the importance of paced instruction, personalised support, and reducing practical barriers \citep{xu2025lifelong}. These studies address different populations and outcomes, and should not be taken as evidence of uniform benefits across age groups.

Existing research examines demographic differences within particular educational settings, including socioeconomic disparities in school students' digital-media access and use \citep{KASTORFF2025105409}. Our surveys describe reported use and perceptions across age groups and countries without directly measuring occupational differences.

\subsection{Age Differences in Digital Engagement}

\citet{ng2012digitalnatives} found that undergraduate students could develop digital literacy through supported use of unfamiliar educational technologies. This study did not establish an age-related gradient in digital engagement; broader claims about a uniform generation of digitally proficient young learners have also been challenged \citep{bennett2008digital}. In a survey of 9,395 students and 964 teachers in Iran, \citet{akbari2025crosssectional} compared preferences for face-to-face, blended, and online education and found that both groups generally preferred face-to-face education. Our study instead examines reported use of named platforms across age groups. The resulting comparisons concern platform use and self-reported learning activities, rather than gains in skills or learning outcomes.

\subsection{Large Language Models and AI in Education}

Large language models (LLMs), including GPT models, can support dialogue and content generation in educational settings. \citet{ennion2025large} conceptualise how LLM chatbots can influence learning behaviours such as resilience, adaptive learning, and challenge-seeking. A meta-analysis of 35 experimental studies published between 2022 and 2024, involving 4,193 participants, reported a moderate positive pooled effect of ChatGPT on student learning outcomes ($g=0.670$); educational level did not significantly moderate this effect \citep{wu2026chatgpt}. Evidence of benefits during AI-assisted tasks should also be distinguished from unaided learning: a field experiment in high-school mathematics found poorer subsequent unaided performance with an unrestricted AI interface, an effect mitigated by a tutor designed with learning safeguards \citep{bastani2025guardrails}. \citet{yan2024practical} emphasise ethical concerns, including bias, dependency, data privacy, and accuracy, underlining the need for responsible integration into educational systems. Long-term pedagogical implications remain underexplored, especially in adult and lifelong learning contexts where structured guidance and goal alignment are critical \citep{beale2025llmstate}. Evaluations should therefore consider both assisted performance and what learners can do without AI support.

Recent studies also examine perceptions of AI and its institutional uses. In a comparative mixed-methods survey of 350 students and faculty, \citet{bhardwaj2026reimagining} found that nearly half of respondents viewed AI- and LLM-based systems as more effective than traditional instruction for comprehension and retention, while traditional methods remained valued for mentorship and ethical development, pointing toward hybrid models. At the institutional level, \citet{bhardwaj2025rpa} review how robotic process automation is being used to streamline administrative and student-support processes in higher education. These studies examine perceived educational benefits and institutional automation; our surveys focus on learners' reported use and preferences.

\subsection{Gamification for Motivation}

We distinguish gamification, which introduces game-design elements into activities that are not themselves games, from the use of complete serious games for learning \citep{deterding2011gamification}.

Systematic reviews have examined how gamification relates to engagement and motivation in school education. For instance, \citet{ruiz2024impact} analysed 90 interventions and found that gamified designs effectively promoted student engagement at primary and secondary school levels. Similarly, \citet{ratinho2023role} reported short-term engagement benefits but noted that motivation may decline with prolonged exposure to gamification. \citet{nylenGameThinking} reported a pooled effect of 0.99 (95\% CI [0.53, 1.44]) on academic achievement for game-thinking interventions in nursing education, an umbrella term covering both serious games and gamification. A separate meta-analysis of gamification reported positive pooled effects on cognitive, motivational, and behavioural outcomes, although motivational and behavioural effects were less stable in analyses restricted to studies with greater methodological rigour \citep{sailer2020gamification}. In an evaluation of the augmented-reality game PEAR, \citet{su132413912} observed improved sustainability knowledge and changes in some self-reported attitudes and behavioural intentions among participants who completed the study. \citet{salderellan2025gamification} reviewed 19 studies in secondary education and reported motivational benefits, including improvements in student agency and the learning environment, alongside mixed findings for longer-term outcomes.

Our surveys describe the perceived motivational value of gamification across age groups. They do not test whether gamification improves retention, whether it is equally effective across populations, or whether demographic personalisation improves learning outcomes.

\subsection{Digital Learning for Workforce Training}

Digital learning can support professional development, but the present study examines reported use and feature preferences rather than workplace performance or training effectiveness. Its relevance to workforce training is therefore exploratory. Research across six massive open online courses (MOOCs) found that goal setting and strategic planning predicted attainment of learners' personal course goals, highlighting learning strategies as another possible explanation for engagement differences \citep{kizilcec2017srl}. The present surveys did not measure these self-regulated learning strategies.

\subsection{Relevance to Present Study}

This study provides an exploratory description of platform use, valued learning features, and perceptions of AI tools among respondents recruited through social networks. It complements existing research on digital literacy and learning support, while its convenience samples limit generalisation beyond the respondents.

We address the following research questions:
\begin{itemize}
    \item \textbf{RQ1:} How do people with diverse demographics engage with different digital learning platforms?
    \item \textbf{RQ2:} Which AI tools and features do respondents report using and valuing across age groups?
    \item \textbf{RQ3:} Which platform features and engagement elements do learners report as most important for digital learning?
\end{itemize}

\section{Methodology}\label{sec:methodology}

\subsection{Survey Participants and Distribution Methods}

We distributed two online surveys about digital-learning use and perceptions. Recruitment used the authors' social networks, primarily contacts in India, Singapore, the UK, and the USA. The surveys recorded age group, gender, and reported country, but did not directly measure occupation or career stage. Participation was voluntary. The study was reviewed and approved by the SUTD Institutional Review Board (IRB) (approval number IRB-24-657), and all procedures were carried out in accordance with the principles of the 1964 Declaration of Helsinki and its later amendments.

We conducted the study using two distinct survey instruments, referred to as Survey~A and Survey~B. Both questionnaires offered the age categories 15--20, 21--30, 31--40, and over 40 years, and the gender options male, female, prefer not to share, and other. The target population was learners engaged in or interested in digital learning across the lifespan; the 15--20 band was included to capture learners at the transition from secondary to tertiary or self-directed learning, so that patterns could be described from late schooling to mid-career and beyond. The IRB-approved protocol set a minimum age of 15 years, so the 15--20 band may include participants aged 15 to 17. The surveys were distributed through Instagram, WhatsApp, and LinkedIn, and responses were collected using Google Forms.

Survey~A recruited 119 participants, of whom 118 provided usable responses (n=118); one incomplete response was excluded. Survey~B contained 81 responses (n=81). The surveys were distributed as separate anonymous forms without a linking identifier, so overlap between respondents cannot be determined. We therefore analysed the surveys separately and did not estimate within-person changes between Survey~A and Survey~B. Survey~A examined reported platform use, recalled relevance, and valued learning features; Survey~B examined reported AI use and perceptions. In Survey~A, country was coded from the free-text answer to the question asking where respondents were from (QA.2); Survey~B asked for country of residence directly (QB.3, free text). In Survey~A, respondents' reported countries were India (58), Singapore (30), the UK (16), and the USA (13), with one unspecified. In Survey~B, they were Singapore (48), India (19), the USA (6), the United Arab Emirates (2), China (2), and Canada, Malaysia, Sri Lanka, and the UK (one each). Survey~A included 65 women, 52 men, and one respondent who preferred not to share their gender; Survey~B included 50 women, 30 men, and one respondent who preferred not to share. Age distributions are shown in Figure~\ref{fig:age-dist}. Survey~A responses were collected between April and December 2024 and Survey~B between March and May 2025. Respondents were recruited by convenience sampling through the authors' social networks. The anonymised datasets, analysis code, and figure code are archived at Zenodo under restricted access and are available on request \citep{puri2026data}.

\subsection{Survey Design and Focus Areas}

Both surveys combined closed-ended questions and optional free-text fields to capture reported digital-learning use and perceptions. Question formats included multiple-choice, checkbox, and Likert-scale items, complemented by free-text fields under an ``Other'' option that allowed respondents to report answers not covered by the predefined categories. The questionnaire supplied a definition of gamification before asking respondents about their awareness of it and its perceived motivational value. Survey~A asked about reported use of entertainment and learning platforms, learning activities, and valued features. Survey~B asked about uses of AI tools and features respondents liked or disliked. These items measured reported experiences and perceptions rather than the effects of particular features on learning. The questionnaires were developed by the authors for this exploratory study and were not formally validated; each construct was measured with a single item, and Likert ratings are treated as ordinal. The items and response options are listed in Appendices~\ref{appendix:survey1} and~\ref{appendix:survey2}.

Three items required coding before analysis. For QA.14 (typical learning-session length), the seven recorded options were grouped into 0--15 minutes (n=21), 15--30 minutes (n=33), over 30 minutes (n=32), and ``until the course is finished'' (n=25); the last is not an ordered duration and is treated as a separate category. Seven responses (``while I feel I have upskilled, 0--60 min'') spanned all categories and were not assigned. For QA.8 (mobile gaming), the six frequency-by-duration options were grouped by daily duration into up to 1 hour (n=88) and over 1 hour per day (n=27); three responses recorded under two non-standard options that are not part of the listed questionnaire (``0--5 hours per week'', n=2; ``0--5 hours per day'', n=1) could not be assigned. Unassigned responses were omitted from Figure~\ref{fig:motivation-bins-agegroup} only and retained in all other analyses. QA.11 (life skills) was addressed to respondents who answered `Yes' to QA.10, but the form had no conditional branching, so it was shown to everyone; 19 respondents who selected life skills without answering `Yes' to QA.10 were excluded from the life-skill analysis, giving a base of 85.

Statistical analyses were conducted in Python (pandas, SciPy, statsmodels). Percentages are calculated over respondents who answered the item, and item non-response is reported where it occurs. For multi-select items, only the predefined options were counted; free-text ``Other'' entries are reported separately where shown and were not analysed qualitatively. For the paired pre- and post-pandemic ratings (QA.5, QA.6; n=118) we used the Wilcoxon signed-rank test \citep{wilcoxon1945}, discarding zero differences (Wilcoxon's method) and using the normal approximation without continuity correction. We report $W$ as the smaller of the two signed-rank sums and the effect size $r=|Z|/\sqrt{N}$, where $N$ is the number of non-zero differences. Differences in the pre--post change and in gamification-motivation ratings (QA.17) across age groups, countries, or session-length categories were tested with Kruskal--Wallis tests, and between women and men or between the two gaming-time categories with Mann--Whitney $U$ tests. The association between selecting ease of use (QA.18) and age group was tested with a chi-square test of independence; because two of its eight cells had expected counts below five, its $p$-value was obtained by Monte Carlo permutation (20,000 permutations). We report 95\% Wilson confidence intervals for headline proportions and $t$-based 95\% confidence intervals for means in figures. The eight tests are exploratory and were not adjusted for multiple comparisons; given the modest, uneven subgroup sizes, inferential results are interpreted cautiously. The archived analysis and figure code reproduce the statistical tests, headline proportions, and figures reported here \citep{puri2026data}.

\subsection{Use of AI Tools}

Generative AI tools were used during the preparation of this manuscript. ChatGPT (OpenAI) and Claude (Anthropic) assisted with language editing. Claude also assisted in writing the Python analysis and plotting code and in drafting textual revisions and the response to reviewers, and an AI-assisted review tool (Astra) was used for a consistency review of the revised manuscript that suggested textual revisions and additional references. AI tools were not used to design the study or to collect data. The authors specified the analyses, checked the code and its outputs against the data, verified every reference against its source, and take full responsibility for the integrity and accuracy of the manuscript.

\section{Results}\label{sec:results}

The following sections describe the samples, reported platform use, and valued learning features in the two surveys.

\subsection{Age Groups}

\begin{figure}[htbp]
    \centering
    \includegraphics[width=0.85\textwidth]{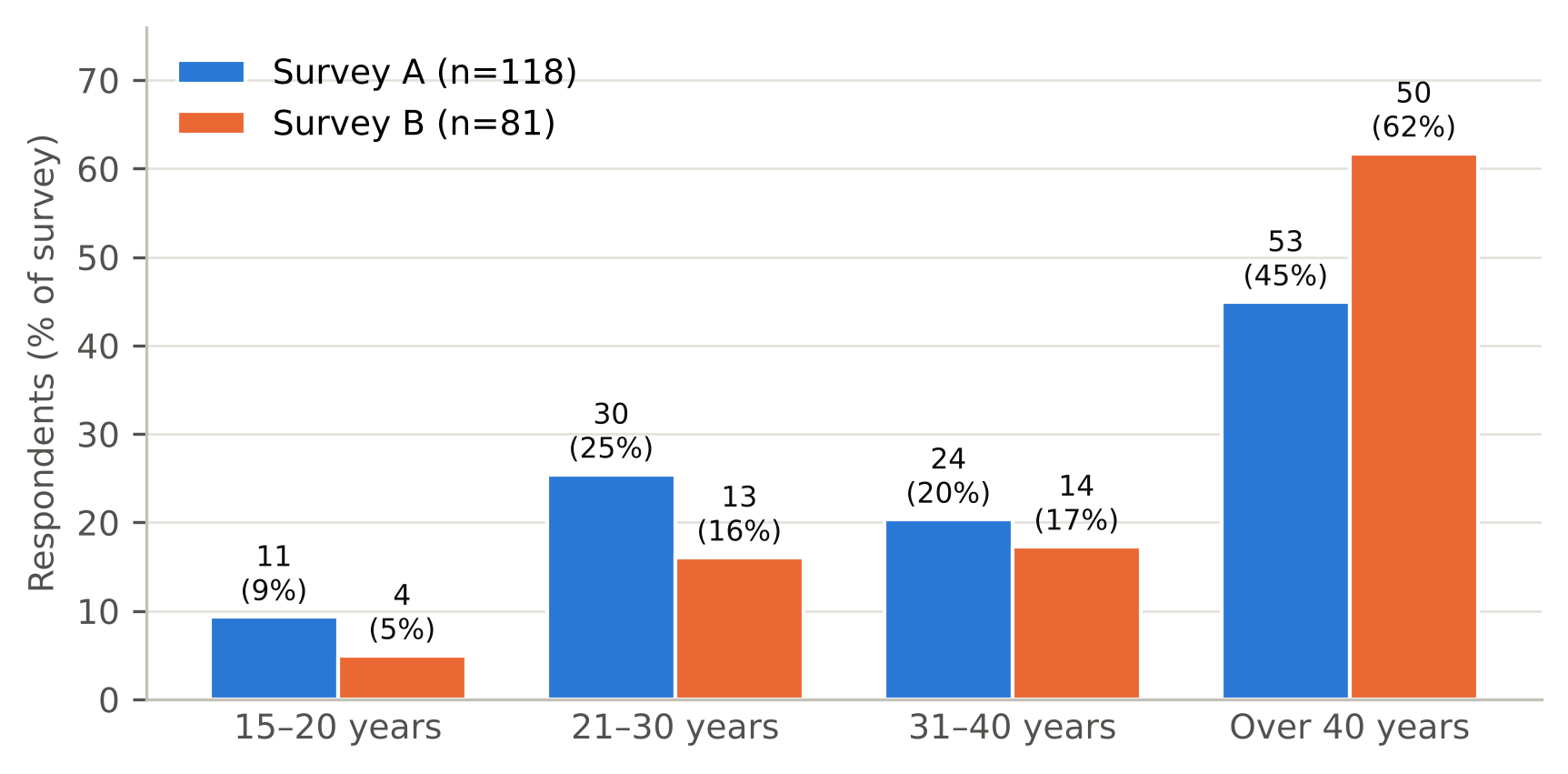}
    \caption{Age distribution of respondents in Survey~A (n=118) and Survey~B (n=81). Bars show the percentage of each survey's respondents; labels give counts.}
    \label{fig:age-dist}
\end{figure}

We examined digital-learning experiences across the age groups represented in the two surveys. Respondents over 40 formed the largest age group in both samples (45\% of Survey~A and 62\% of Survey~B). Figure~\ref{fig:age-dist} shows the age distribution of the respondents in Survey~A and Survey~B.

\subsection{Recalled Relevance and Reported Use of Digital Learning}

\begin{figure}[htbp]
\centering
\includegraphics[width=0.85\textwidth]{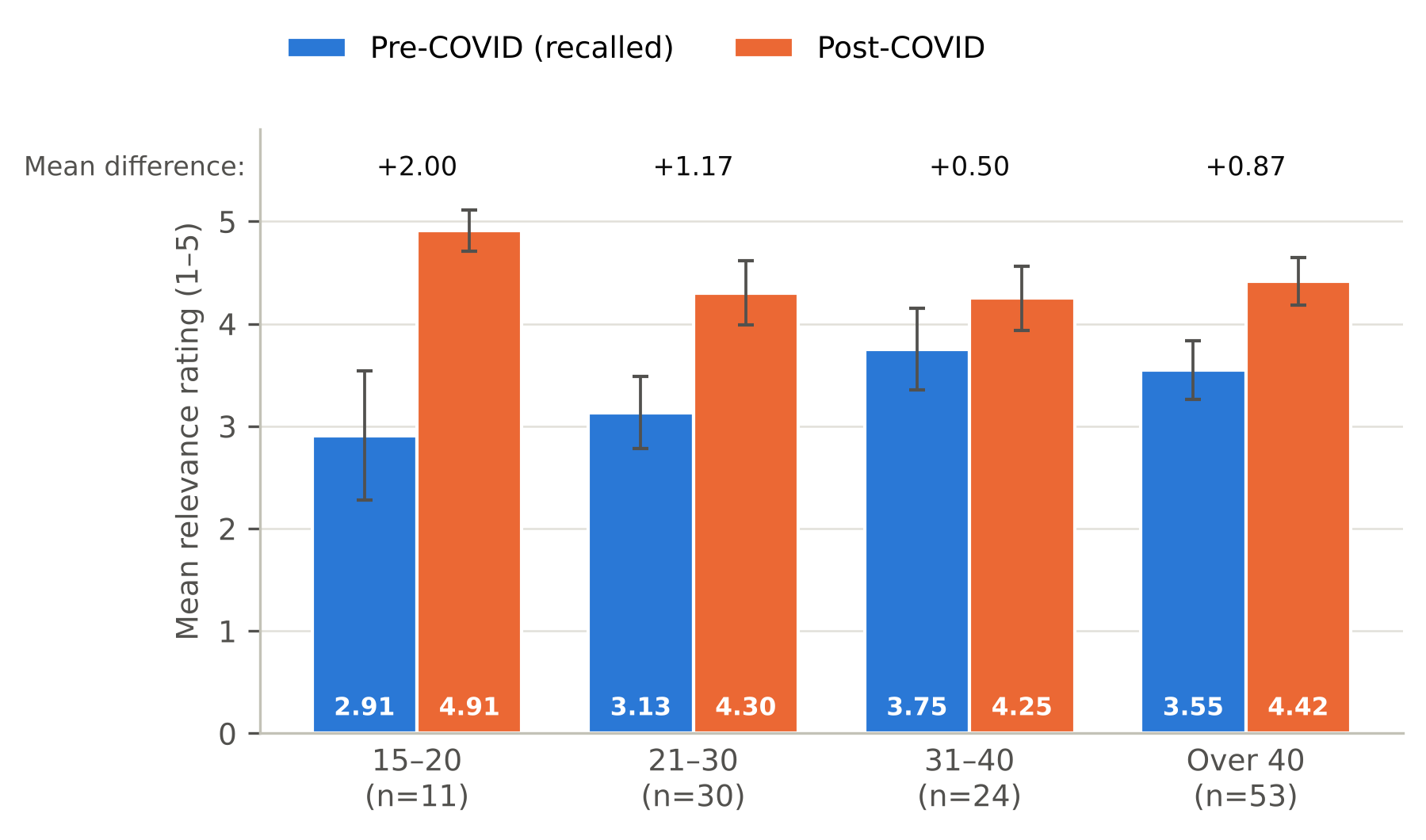}
\caption{Mean recalled relevance of digital learning before and after COVID-19, by age group (Survey~A, QA.5--QA.6; 1 = not relevant at all, 5 = extremely relevant). Error bars show 95\% confidence intervals; values above each pair are mean differences (post minus pre).}
\label{fig:pre-post-rel}
\end{figure}

Survey~A asked respondents to rate the relevance of digital learning to them before and after COVID-19 on a five-point scale (QA.5, QA.6). Figure~\ref{fig:pre-post-rel} shows the mean recalled ratings for each period by age group. The youngest age group (15--20 years) showed the largest descriptive increase in recalled relevance. When asked whether they used digital learning materials more often than before COVID-19, 100 of 118 respondents (84.7\%, 95\% CI 77.2--90.1\%) selected `Yes'; the other options were `No' and `Same as pre-COVID'. Three further responses hold values not offered in the final form (`Increased', n=2; `Just the same', n=1) and were not counted as `Yes'.

Respondents retrospectively rated digital learning as more relevant after the pandemic than before: 69 of the 118 paired ratings increased, 6 decreased, and 43 were unchanged. A Wilcoxon signed-rank test on these paired retrospective ratings indicated that the difference in recalled relevance was statistically significant ($W = 131.0$, $Z=-6.97$, $p < 0.001$; effect size $r = 0.80$ with $N=75$ non-zero differences, a large effect). Because both ratings were provided in a single sitting, this reflects a difference in \emph{recalled} relevance rather than a measured change over time (see Section~\ref{sec:conclusion}). The magnitude of the increase differed significantly across age groups (Kruskal--Wallis $H(3) = 14.02$, $p = 0.003$), with the youngest group (15--20) showing the largest mean increase (2.00 points, n=11) and the 31--40 group the smallest (0.50 points, n=24); this omnibus test does not identify which pairs of groups differ. The increase did not differ significantly by gender (Mann--Whitney $U = 1834$, $p = 0.41$; 65 women and 52 men) or country (Kruskal--Wallis $H(3) = 1.04$, $p = 0.79$).

\subsubsection{Recalled Relevance by Gender and Country}

Figure~\ref{fig:trend_ageGroups_Gender_Region} presents mean recalled relevance by reported country and gender:

\begin{figure}[htbp]
\centering
\includegraphics[width=0.85\textwidth]{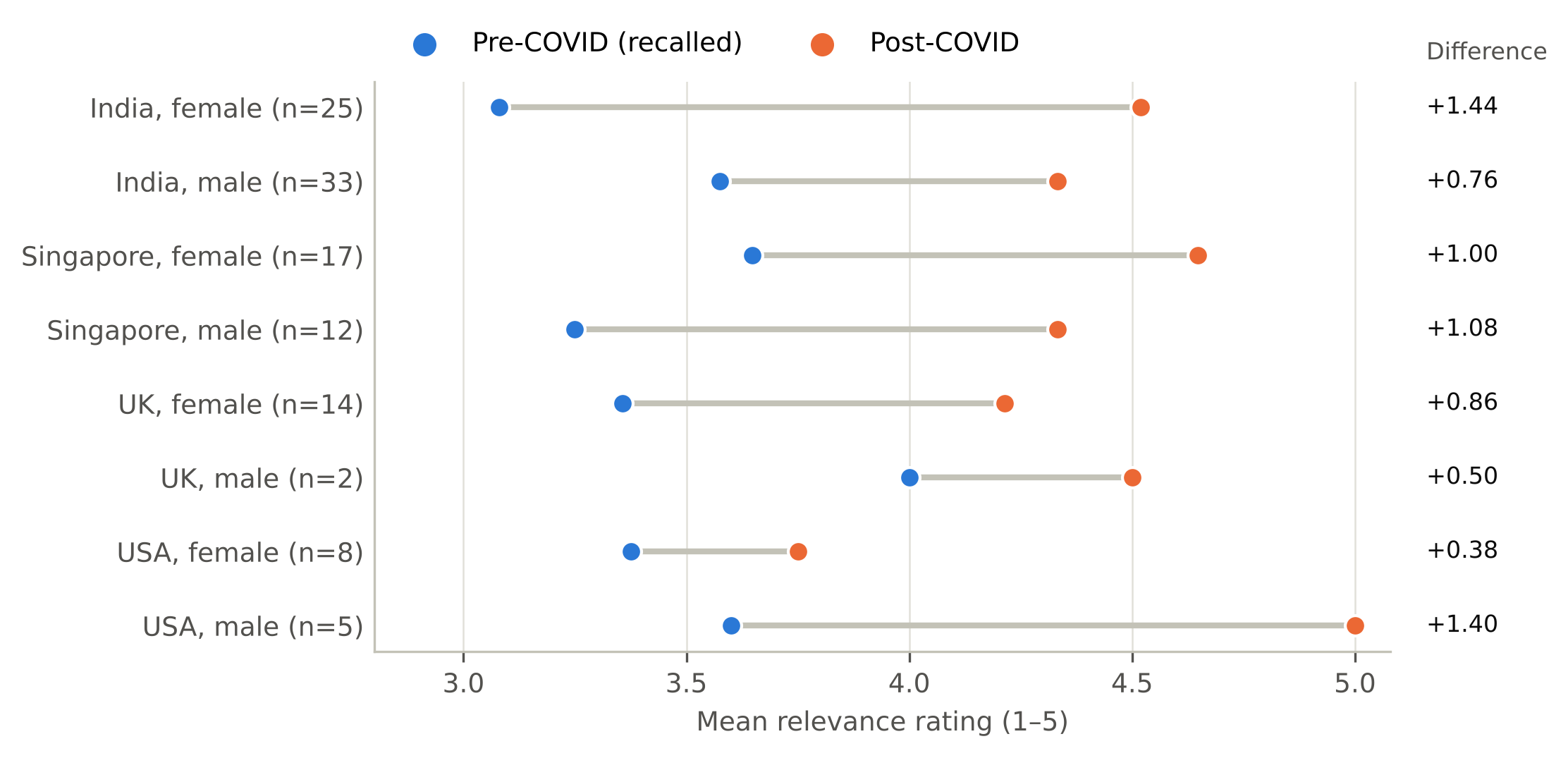}
\caption{Mean recalled relevance of digital learning before and after COVID-19 by reported country and gender (Survey~A; 116 respondents who reported a country and identified as female or male). Values on the right are mean differences (post minus pre). Cells with fewer than ten respondents are descriptive only.}
\label{fig:trend_ageGroups_Gender_Region}
\end{figure}

\begin{itemize}
    \item Respondents in India and Singapore reported increases in recalled digital-learning relevance among both women and men. Women in India reported a mean increase of 1.44 points (n=25), compared with 0.76 points among men in India (n=33).
    \item Male respondents in the USA reported a large shift, from a moderate pre-pandemic mean (3.6) to the scale maximum post-COVID (5.0; n=5); a cell mean of exactly 5.0 means every respondent in this small cell of five selected the maximum, so it should be read with caution. In the UK, women reported a mean increase of 0.86 points (n=14); the UK male cell contained only two respondents. Several country$\times$gender cells are small, and comparisons based on fewer than ten respondents are reported descriptively only.
\end{itemize}

\subsubsection{Reported Platform Use by Age and Country}
\label{subsec:pref-plat-by-age}

\begin{figure}[htbp]
    \centering
    \begin{subfigure}[b]{0.85\textwidth}
        \centering
        \includegraphics[width=\textwidth]{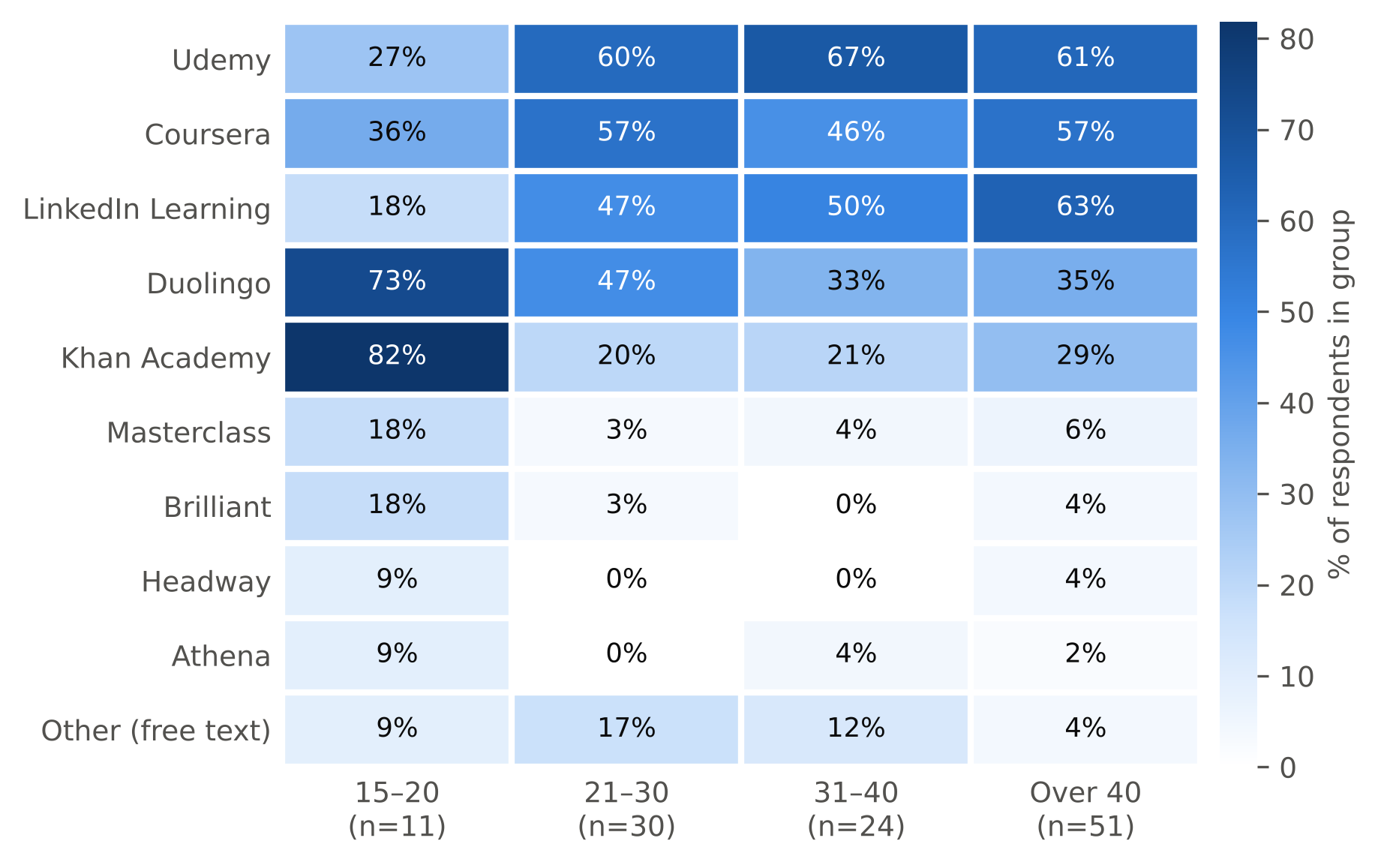}
        \caption{By age group}
        \label{fig:pref-learn-plat}
    \end{subfigure}
    \vspace{0.4cm}
    \begin{subfigure}[b]{0.85\textwidth}
        \centering
        \includegraphics[width=\textwidth]{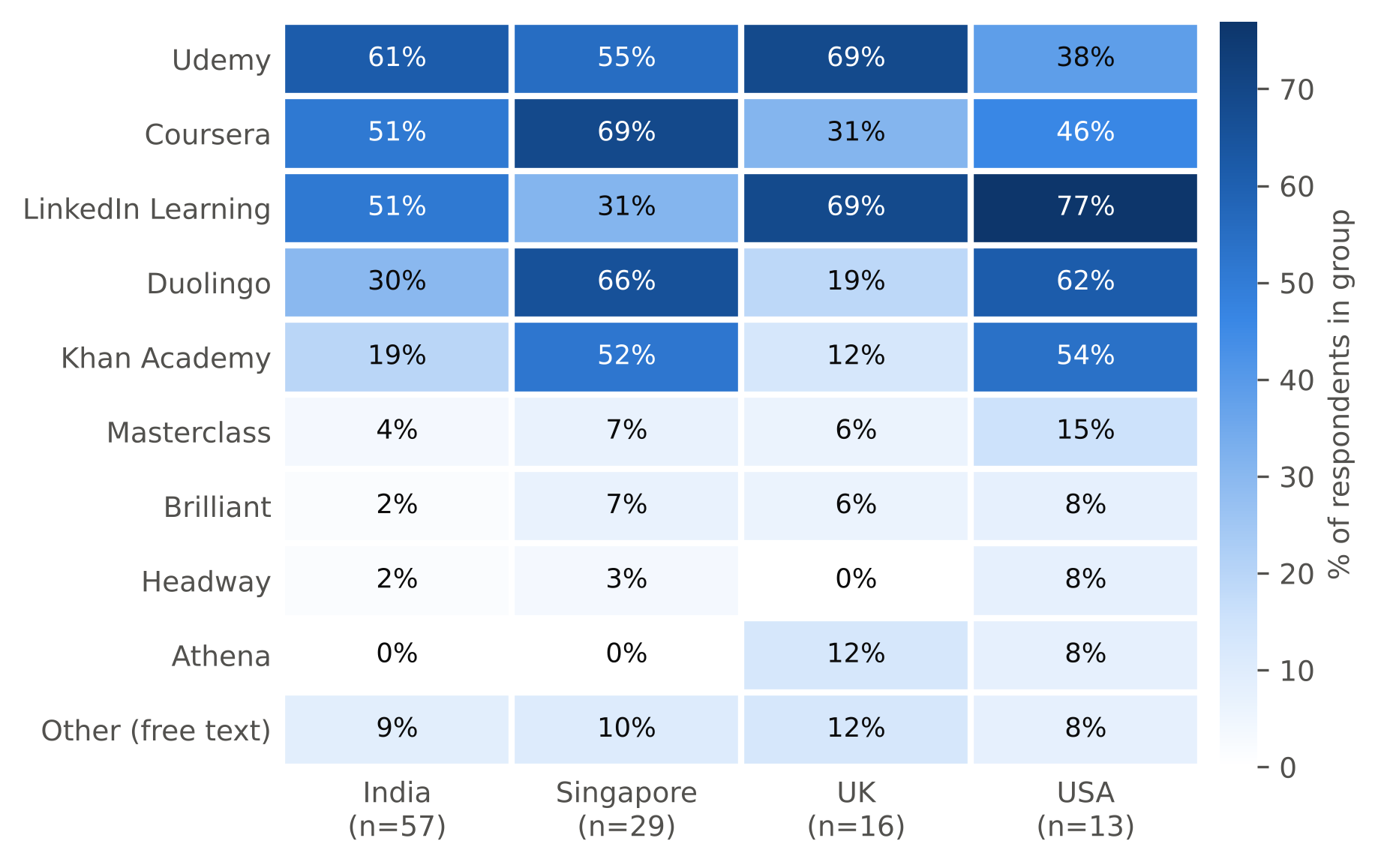}
        \caption{By reported country}
        \label{fig:pref-region}
    \end{subfigure}
    \caption{Percentage of respondents in each age group (a) and reported country (b) who reported using each learning platform (Survey~A, QA.9; 116 respondents answered; more than one platform could be selected, so columns do not sum to 100\%). ``Other (free text)'' counts respondents who named at least one platform not listed in the question. One respondent with an unspecified country is omitted from (b).}
    \label{fig:combined-platforms}
\end{figure}

Reported platform use varied descriptively across age groups (Figure~\ref{fig:pref-learn-plat}). Khan Academy (9 of 11; 82\%) and Duolingo (8 of 11; 73\%) were most frequently reported by respondents aged 15--20, whereas Udemy, Coursera, and LinkedIn Learning were more frequently reported by older groups (for example, LinkedIn Learning by 63\% of respondents over 40, compared with 18\% of those aged 15--20). Country patterns were not uniform: for example, Duolingo was reported more frequently than Udemy in Singapore (66\% vs 55\%) and the USA (62\% vs 38\%) (Figure~\ref{fig:pref-region}). These differences describe platform use within this sample and do not establish platform preferences or learning effectiveness.

\subsection{Mapping Demographics to Platforms and Skills}

\subsubsection{Learning After Graduation and Life-Skill Domains}

\begin{figure}[htbp]
\centering
\includegraphics[width=\textwidth]{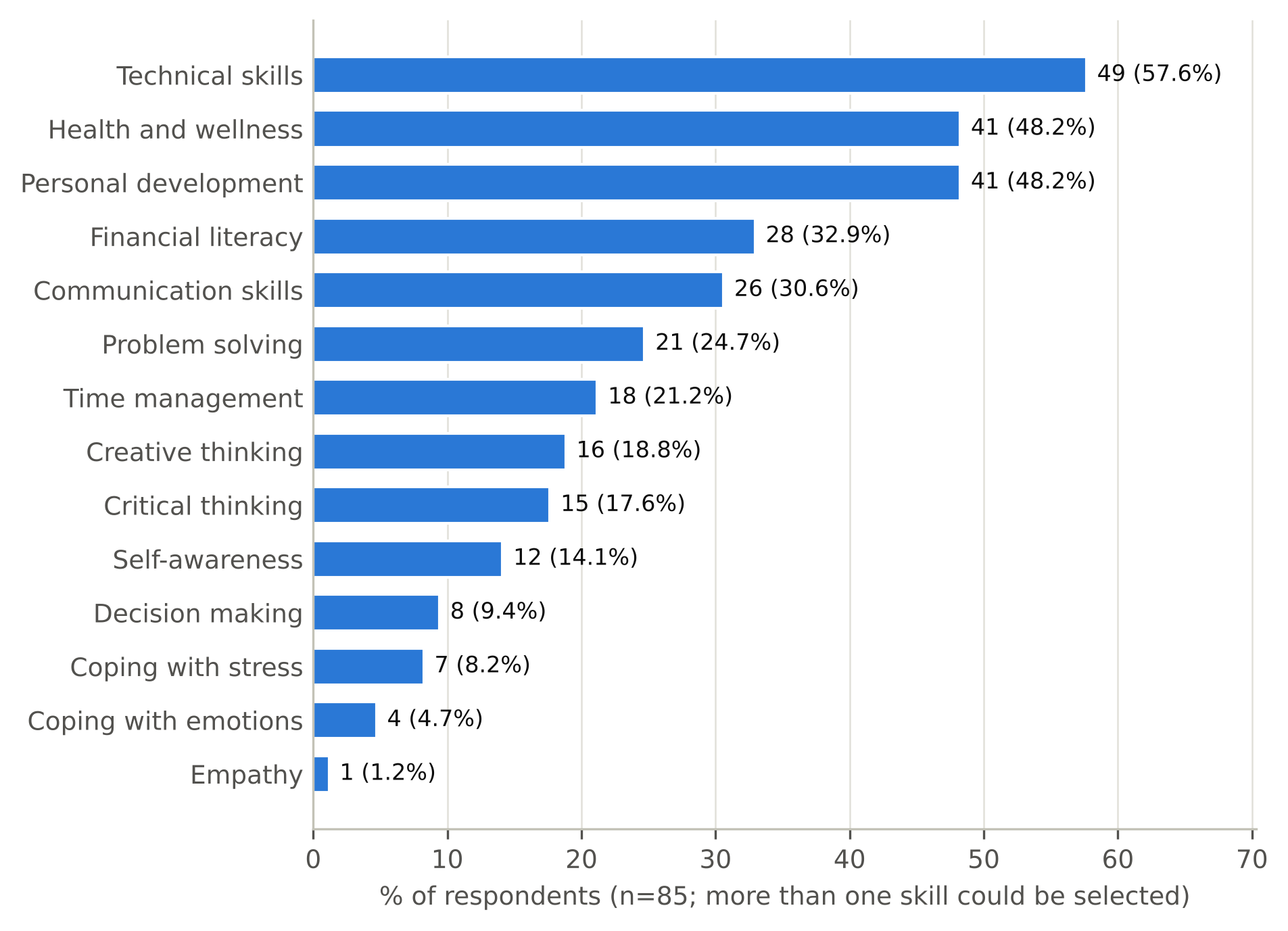}
\caption{Life-skill domains reported in QA.11 by respondents who answered `Yes' to QA.10 (n=85; more than one domain could be selected).}
\label{fig:life-skills}
\end{figure}

We also asked whether respondents had used online platforms for lifelong learning after graduating (QA.13). Of the 75 respondents aged 31 and above who answered this item (two did not), 59 (79\%) reported doing so. Among the 85 respondents who used e-learning for life skills (QA.10) and selected at least one listed domain (QA.11), technical skills (49; 57.6\%), health and wellness (41; 48.2\%), and personal development (41; 48.2\%) were reported most often (Figure~\ref{fig:life-skills}). Descriptively, technical skills were selected by 19 of 27 respondents aged 21--30 (70\%), 7 of 16 aged 31--40 (44\%), and 22 of 39 over 40 (56\%); only three respondents aged 15--20 were in this base. Technical skills were selected by 25 of 37 men (68\%) and 23 of 47 women (49\%), and health and wellness by 13 of 37 men (35\%) and 28 of 47 women (60\%). The surveys did not record occupation, so these patterns cannot be described by professional background.

\subsubsection{Co-reported Platform Use and Skill Domains}

\begin{figure}[htbp]
\centering
\includegraphics[width=0.85\textwidth]{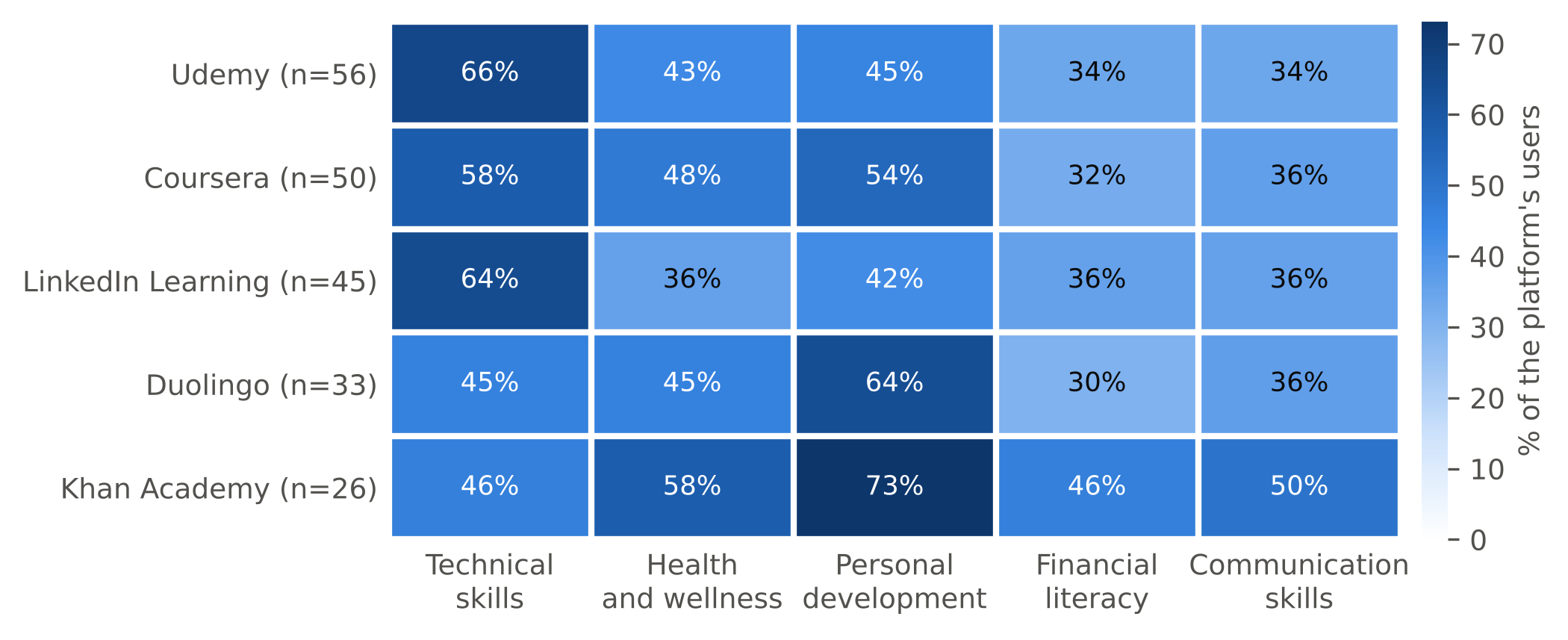}
\caption{Percentage of each platform's users who reported each of the five most frequent life-skill domains (respondents in the life-skill base, n=85; n per platform in brackets). Platforms (QA.9) and skills (QA.11) were reported in separate multi-select questions, so cells are co-selections rather than platform-specific learning.}
\label{fig:skill-platform-heatmap}
\end{figure}

Figure~\ref{fig:skill-platform-heatmap} shows, for each of the five most frequently reported platforms, the percentage of its users in the life-skill base who also reported each life-skill domain. Technical skills were the most common domain among Udemy (66\%) and LinkedIn Learning (64\%) users, whereas personal development was the most common among Duolingo (64\%) and Khan Academy (73\%) users. Because platforms and skills were reported in separate multi-select questions, these are co-selections: they do not show that respondents used a given platform to learn a given skill. We therefore interpret the figure descriptively and do not apply a single chi-square test to the overlapping multi-response table.

\subsection{Valued Features and Perceived Motivation}

\begin{figure}[htbp]
\centering
\includegraphics[width=\textwidth]{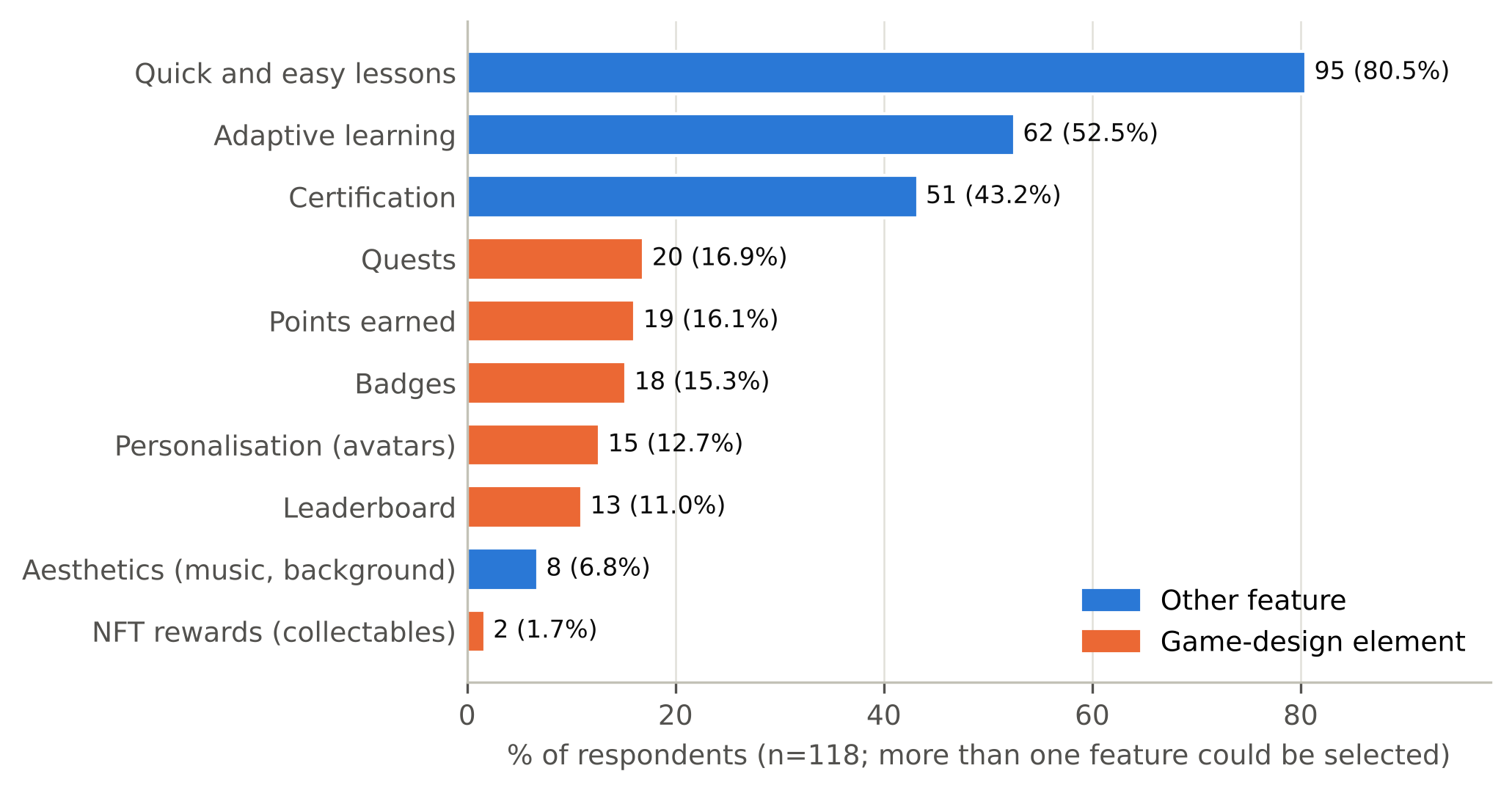}
\caption{Features respondents liked about learning with apps (Survey~A, QA.15; n=118; more than one feature could be selected). Game-design elements are shown in orange.}
\label{fig:liked-feature-apps}
\end{figure}

When asked what they liked most about learning with apps (QA.15; answered by all 118 respondents, more than one feature could be selected), respondents most often selected quick and easy lessons (95; 80.5\%), adaptive learning (62; 52.5\%), and certification (51; 43.2\%, 95\% CI 34.6--52.2\%) (Figure~\ref{fig:liked-feature-apps}). Each game-design element (quests, points, badges, avatars, leaderboards, and collectable rewards) was selected by 17\% of respondents or fewer. We treat certification separately because a course-completion credential is not necessarily a game element. The next subsection examines the perceived motivational value of gamification across age groups.

\subsubsection{Gamification as a Motivating Feature}

Survey~A asked respondents to rate their awareness of gamification and its perceived motivational value on five-point scales (QA.16, QA.17). The highest awareness rating was selected by 37 of 118 respondents (31.4\%), and the second highest by 23 (19.5\%). For perceived motivational value, 23 (19.5\%) selected the highest rating and 35 (29.7\%) the second highest. These responses describe perceptions rather than demonstrated effects on motivation or learning.

\begin{table}[htbp]
\centering
\caption{Gamification-motivation rating (QA.17; 1--5) by age group (Survey~A)}
\label{tab:motivation-age}
\begin{tabular}{lccc}
\toprule
\textbf{Age Group} & \makecell{\textbf{Mean} \\ \textbf{Motivation Rating}} & \makecell{\textbf{Respondent} \\ \textbf{Count}} & \makecell{\textbf{Standard} \\ \textbf{Deviation (SD)}} \\
\midrule
15 to 20 years & 3.64 & 11 & 0.81\\
21 to 30 years & 3.40 & 30 & 1.28\\
31 to 40 years & 3.12 & 24 & 1.48\\
Over 40 years  & 3.25 & 53 & 1.33\\
\midrule
\textbf{Overall} & \textbf{3.30} & \textbf{118} & \textbf{1.30} \\
\bottomrule
\end{tabular}
\end{table}

Table~\ref{tab:motivation-age} presents mean gamification-motivation ratings by age group. Mean ratings were similar across age groups (3.12 to 3.64 on a five-point scale), with the youngest group (15--20) highest; however, a Kruskal--Wallis test found no statistically significant difference across the four age groups ($H(3) = 0.98$, $p = 0.81$). The apparent variation is therefore descriptive only, and the small 15--20 group (n=11) further limits precision. A non-significant result does not establish equivalence between age groups.

\begin{figure}[htbp]
    \centering
    \begin{subfigure}[b]{0.55\textwidth}
        \centering
        \includegraphics[width=\textwidth]{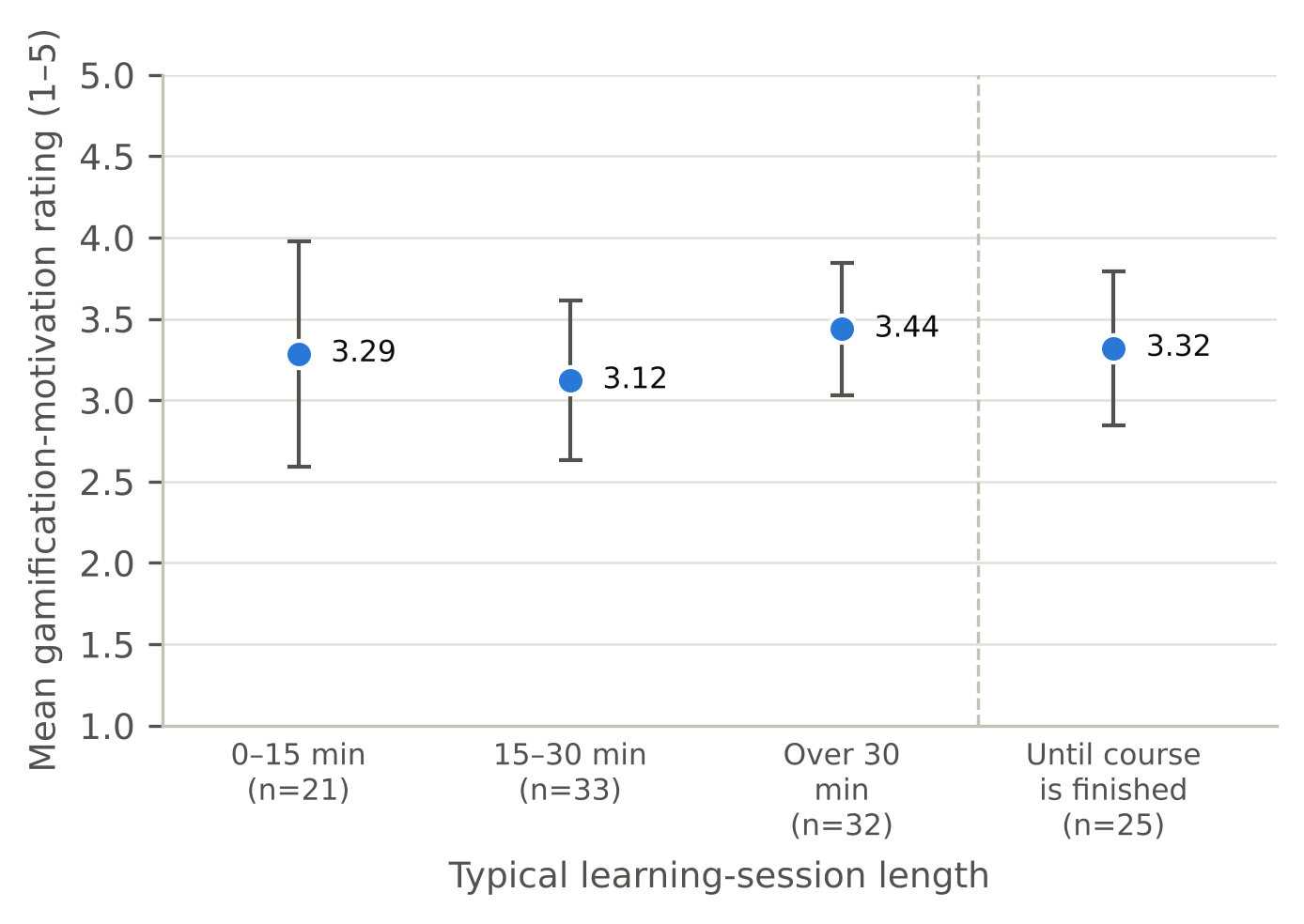}
        \caption{By typical learning-session length}
        \label{fig:time_spent_learning}
    \end{subfigure}
    \hspace*{0.02\textwidth}
    \begin{subfigure}[b]{0.38\textwidth}
        \centering
        \includegraphics[width=\textwidth]{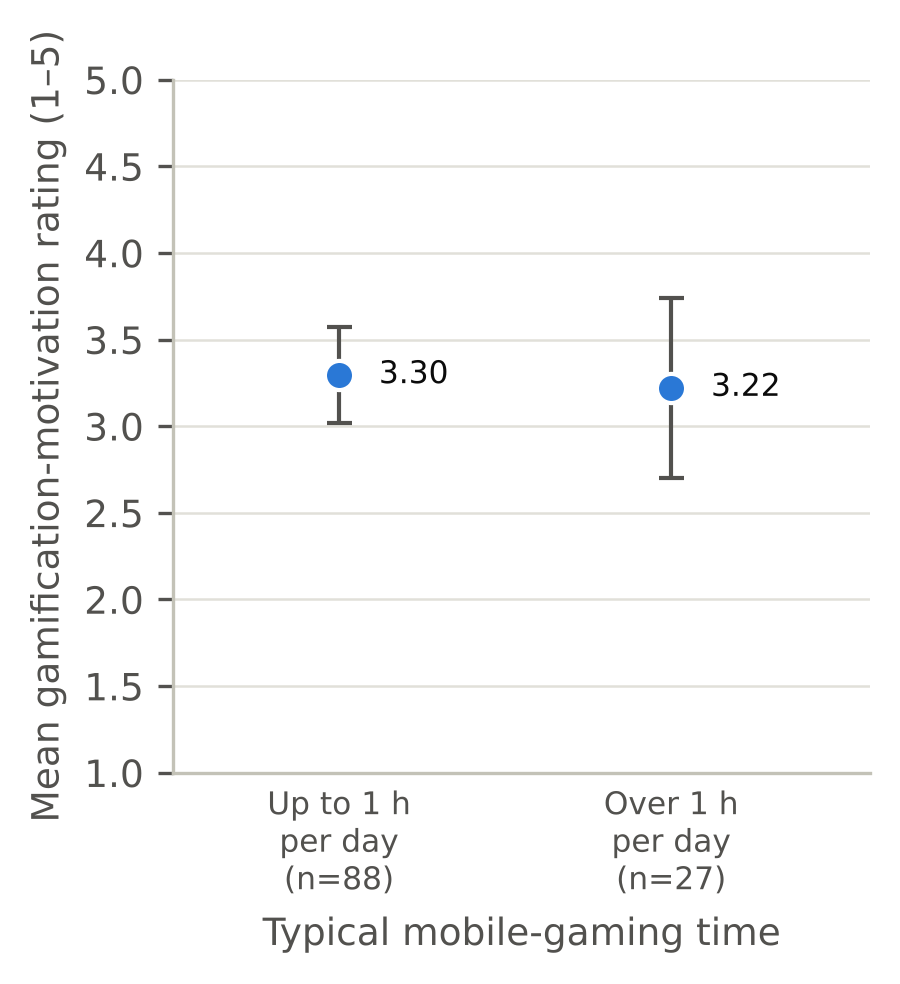}
        \caption{By typical mobile-gaming time}
        \label{fig:time_spent_playing}
    \end{subfigure}
    \caption{Mean gamification-motivation rating (QA.17; 1--5) by typical learning-session length (a; QA.14) and typical mobile-gaming time (b; QA.8), pooled across age groups. Error bars show 95\% confidence intervals. ``Until course is finished'' is not an ordered duration and is shown separately; responses that could not be assigned to a category are omitted (Section~3.2).}
    \label{fig:motivation-bins-agegroup}
\end{figure}

Figure~\ref{fig:motivation-bins-agegroup} shows mean gamification-motivation ratings by typical learning-session length and typical mobile-gaming time. Ratings are pooled across age groups because the age-by-category cells were small (for example, two or three respondents per session-length category in the 15--20 group). Mean ratings were similar across the three ordered session lengths (Kruskal--Wallis $H(2) = 0.55$, $p = 0.76$) and between respondents who reported up to 1 hour and over 1 hour of mobile gaming per day (Mann--Whitney $U = 1235$, $p = 0.75$). These cross-sectional comparisons do not show how motivation changes within an individual's learning session.

\subsubsection{Features Valued for Learning with Apps}

\begin{figure}[htbp]
\centering
\includegraphics[width=0.85\textwidth]{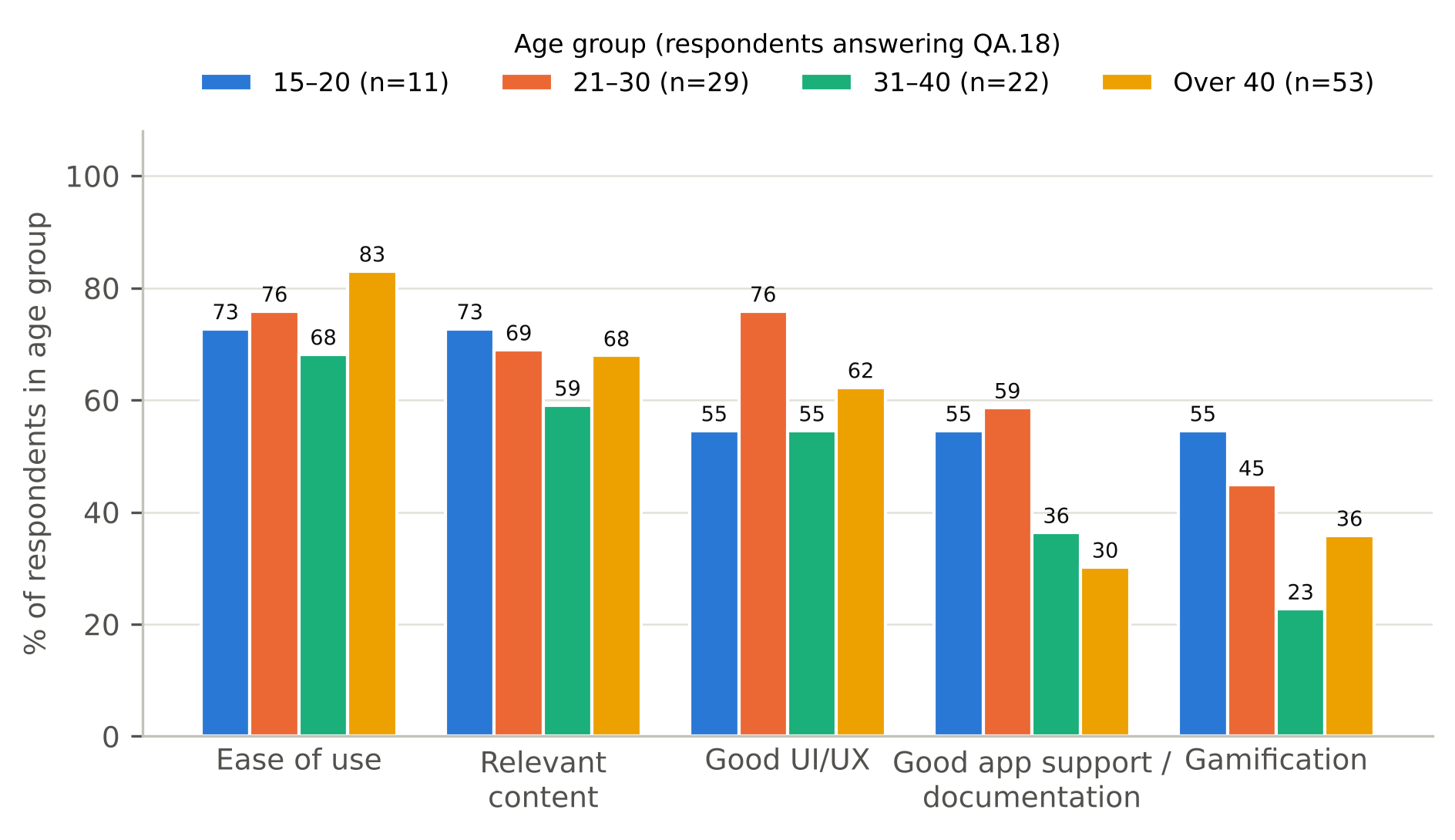}
\caption{Features respondents thought would make learning with apps more conducive (Survey~A, QA.18), as a percentage of respondents in each age group who answered the item (n=115; more than one feature could be selected).}
\label{fig:features-agegroup}
\end{figure}

Survey~A asked which features respondents thought would make learning with apps more conducive (QA.18; answered by 115 respondents, more than one feature could be selected). Overall, 89 of 115 respondents (77.4\%, 95\% CI 68.9--84.1\%) selected ease of use, followed by relevant content (77; 67.0\%), good UI/UX (73; 63.5\%), good app support or documentation (47; 40.9\%), and gamification (43; 37.4\%). Ease of use was the most frequently selected feature, or tied for most frequent, in every age group, and gamification was selected less frequently than ease of use and relevant content in every age group (Figure~\ref{fig:features-agegroup}). Selecting ease of use was not associated with age group ($\chi^2(3) = 2.20$, n=115; Monte Carlo permutation $p = 0.54$). These percentages describe preferences within this sample and do not establish the effectiveness of these features in improving learning.

\subsection{Reported Uses and Perceptions of AI Tools}
\label{sec:liked-features-gpt}

\begin{figure}[htbp]
    \centering
    \includegraphics[width=0.85\textwidth]{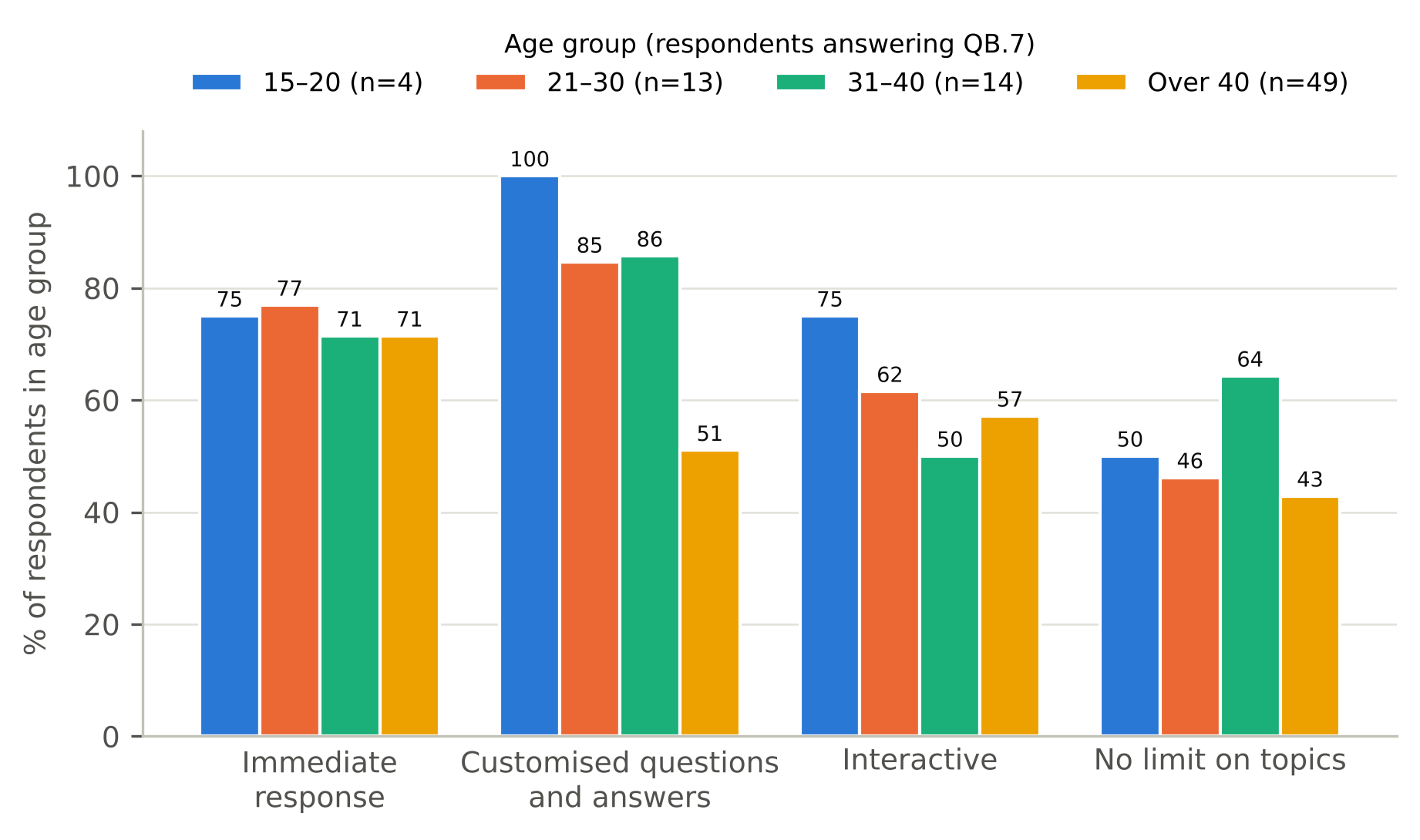}
    \caption{AI features respondents liked (Survey~B, QB.7), as a percentage of respondents in each age group who answered the item (n=80; more than one feature could be selected). The 15--20 group contained four respondents.}
    \label{fig:ai-features-age}
\end{figure}

Survey~B asked about uses of AI tools (QB.5; answered by 80 of 81 respondents, more than one use could be selected). The most commonly reported uses were text generation (56; 70.0\%, 95\% CI 59.2--78.9\%), conceptual understanding (52; 65.0\%), and research (42; 52.5\%). Respondents also selected their favourite AI tools and the features they liked (QB.7; answered by 80). Immediate responses (58), customised questions and answers (52), interactive content (46), and no limit on topics (38) were the features most often liked (Figure~\ref{fig:ai-features-age}).

Figure~\ref{fig:ai-features-age} shows that immediate responses were selected by 71--77\% of respondents in every age group, whereas customised questions and answers were selected less frequently in the over-40 group (51\%) than in the younger groups (85--100\%). Neither immediate responses nor topic variety showed a monotonic increase with age. These comparisons are descriptive, particularly for the 15--20 group, which contained only four respondents.

\begin{figure}[htbp]
\centering
\includegraphics[width=\textwidth]{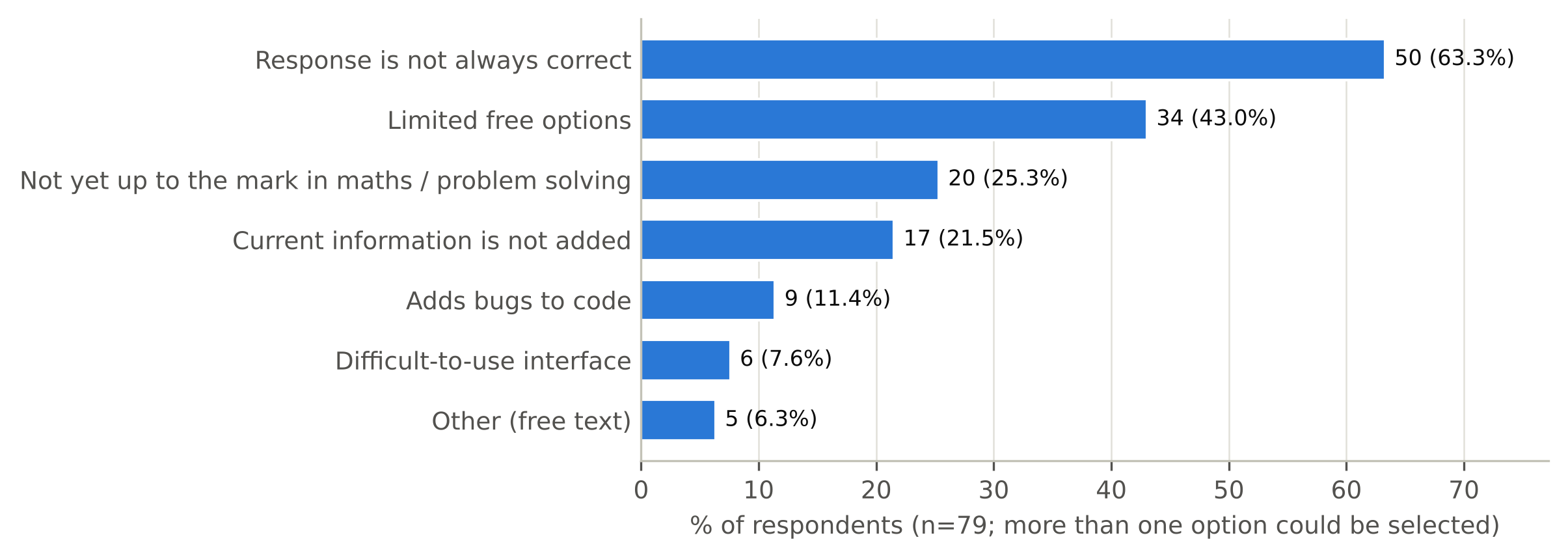}
\caption{Features respondents did not like about learning with AI tools (Survey~B, QB.8; n=79; more than one option could be selected; free-text answers are grouped as Other).}
\label{fig:gpt-dislikes}
\end{figure}

Survey~B also asked what respondents did not like about learning with AI tools (QB.8; a multiple-response item with an optional free-text field, answered by 79 respondents). ``Response is not always correct'' was selected by 50 of 79 respondents (63.3\%, 95\% CI 52.3--73.1\%), followed by limited free options (34; 43.0\%) (Figure~\ref{fig:gpt-dislikes}); one further respondent wrote that they were not sure whether answers would be accurate.

\section{Discussion}\label{sec:discussion}

We discuss respondents' reported platform use, valued learning features, and perceptions of AI tools. Respondents retrospectively reported substantially higher relevance of digital learning after the pandemic, and 84.7\% reported more frequent use, consistent with the pandemic's widely documented role as a catalyst for educational change. This finding concerns recalled relevance and reported use, whereas \citet{zheng2021online} examined course performance during the transition to online teaching in 2020. We discuss demographic patterns in reported platform use (RQ1), uses and perceptions of AI tools (RQ2), and valued learning features (RQ3).

\subsection{How do people with diverse demographics engage with different digital learning platforms?}

Reported platform use differed descriptively across age groups. Respondents aged 15--20 more frequently reported using Khan Academy and Duolingo, while older groups more frequently reported using Udemy, Coursera, and LinkedIn Learning (Figure~\ref{fig:pref-learn-plat}). The survey did not measure platform-specific use before and after the pandemic, employment status, or career stage, so these patterns cannot establish changes in professional learning behaviour.

Figure~\ref{fig:trend_ageGroups_Gender_Region} shows relatively large descriptive increases in recalled relevance among women in India and men in the USA; the latter cell contained only five respondents. In Singapore, the increase was similar for men and women (1.08 and 1.00 points). The increase did not differ significantly by country (Kruskal--Wallis $H(3)=1.04$, $p=0.79$) or gender (Mann--Whitney $U=1834$, $p=0.41$). These comparisons do not establish a basis for country-specific platform recommendations.

The platform--skill cross-tabulation describes co-reported activities within this convenience sample. It may inform hypotheses about platform use and learning goals, but it does not establish global preferences, occupational differences, or the effectiveness of age-specific platform design. Because QA.9 listed named platforms, reported use may partly reflect brand awareness, and the prominence of technical skills may partly reflect the large supply of technical content on these platforms.

\subsection{Which AI tools and features do respondents report using and valuing across age groups?}

AI tools were widely used within the Survey~B sample, with 56 of 80 respondents (70\%) reporting text generation as a use. At the same time, the frequently reported concern about accuracy (50 of 79; 63\%) underscores the value of verifying AI output, for example through hybrid AI--human checking. Respondents valued immediate responses, customised questions and answers, and topic variety. These preferences do not establish that AI improves feedback quality or substitutes for instructor support.

\subsection{Which platform features and engagement elements do learners report as most important for digital learning?}

Survey~A respondents more often selected ease of use and relevant content than gamification as desirable platform features (Figure~\ref{fig:features-agegroup}). Certification was selected by 51 of 118 respondents (43.2\%; Figure~\ref{fig:liked-feature-apps}). However, these responses do not establish why particular platforms are popular, whether certificates improve employment prospects, or which features sustain engagement over time. Mean gamification-motivation ratings did not differ significantly across age groups, session lengths, or gaming times, and the session-length comparisons do not track changes within individuals. The data therefore do not support, although with a small youngest group they cannot rule out, the common expectation that younger learners are more motivated by gamification. The Community of Inquiry framework distinguishes teaching, social, and cognitive presence \citep{Garrison2000}, but the present survey did not measure these constructs. Future longitudinal or experimental studies could test whether the reported feature preferences predict sustained participation or learning outcomes.

\section{Conclusion and Implications}\label{sec:conclusion}

Two separately analysed surveys provide an exploratory account of reported digital-platform use and learning-feature preferences. In Survey~A, respondents recalled digital learning as more relevant after the pandemic, and 100 of 118 respondents (84.7\%) reported more frequent use than before COVID-19. Reported platform use varied descriptively across age groups, but mean gamification-motivation ratings did not differ significantly between them. Ease of use was selected by 89 of 115 respondents (77.4\%), indicating its salience within this sample. In Survey~B, text generation was reported by 56 of 80 respondents (70\%), and 50 of 79 (63\%) selected inaccurate responses as a concern.

These findings suggest usability, relevant content, and reliable AI responses as priorities for further evaluation. They do not establish effects on learning, retention, employment outcomes, or the success of particular platform designs. Social-network recruitment, uneven subgroup sizes, uncertain respondent overlap between surveys, and retrospective ratings limit interpretation. In particular, because the pre- and post-pandemic ratings were given in a single sitting, they are subject to recall bias. Differences between the two recalled periods may also reflect changes in respondents' age, educational stage, or work circumstances, and cannot be attributed to the pandemic or to AI tools. Future research should recruit more broadly, justify sample sizes for its planned comparisons, and use longitudinal or experimental designs to test whether reported preferences predict sustained participation and learning outcomes.


\backmatter

\bmhead{Acknowledgements}
Not applicable.

\section*{Declarations}

\bmhead{Funding}
No funding was received for conducting this study.

\bmhead{Competing Interests}
The authors have no competing interests to declare that are relevant to the content of this article.

\bmhead{Ethical Approval}
The protocol for this study was reviewed and approved by the SUTD Institutional Review Board (IRB) (approval number IRB-24-657). All procedures involving human participants were carried out in accordance with the ethical standards of the institutional review board and with the principles of the 1964 Declaration of Helsinki and its later amendments.

\bmhead{Consent to Participate}
Informed consent to participate was obtained from all participants. Before starting the survey, every participant was presented with an information sheet describing the purpose of the study, that participation was voluntary, that participants could withdraw at any point during the survey by closing the browser, in which case their responses were discarded, and that responses would be anonymous. Participants acknowledged this information and gave consent before proceeding. For participants aged 15 to 17, the IRB approved a waiver of parental consent, and these participants gave their own consent through the same procedure. The surveys did not collect names, contact details, or other direct identifiers, and age was recorded only in bands. The free-text answers naming a school or university (QA.2, QB.2) are not included in the shared dataset; in Survey~A, the answer to QA.2 was used only to code country.

\bmhead{Consent to Publish}
Not applicable. The manuscript does not contain any individual person's identifiable data, images, or personal information.

\bmhead{Availability of Data and Materials}
The anonymised datasets, together with the analysis and figure code that reproduce the reported statistical tests and figures, are deposited in the Zenodo repository under restricted access (\url{https://doi.org/10.5281/zenodo.24156492}) \citep{puri2026data}. The data are not publicly available; they may be shared with other researchers on reasonable request to the corresponding author. Both questionnaires are reproduced in Appendices~\ref{appendix:survey1} and~\ref{appendix:survey2}.


\bibliography{references_v3}

\begin{appendices}

\section{Questionnaire: Survey A}\label{appendix:survey1}

Items and response options are reproduced from the administered Google Form (``Survey to understand online learning behaviours, pre and post Covid, with and without gamification''). Before the questions, participants read an information sheet and ticked a box to give consent. The form had no conditional branching. ``Other'' denotes an optional free-text field.

\begin{enumerate}[noitemsep, topsep=0pt]
    \item[\textsc{QA.1} --] \textbf{You are a} Male; Female; Prefer not to share; Other.
    \item[\textsc{QA.2} --] \textbf{You are from --- (School / University and Country)} Open-ended response.
    \item[\textsc{QA.3} --] \textbf{How young are you?} 15 to 20 years; 21 to 30 years; 31 to 40 years; Over 40 years.
    \item[\textsc{QA.4} --] \textbf{Do you use digital learning materials more often now as compared to pre Covid days?} Yes; No; Same as pre Covid.
    \item[\textsc{QA.5} --] \textbf{How important or relevant was digital learning for you in pre Covid days?} 1 (not relevant at all) to 5 (extremely relevant).
    \item[\textsc{QA.6} --] \textbf{How important or relevant is digital learning for you now, in post Covid days?} 1 (not relevant at all) to 5 (extremely relevant).
    \item[\textsc{QA.7} --] \textbf{What entertainment / gaming apps have you used?} Twisted Tangle; Royal Match; Last War: Survival; Call of Duty; We are warriors!; Township; Candy Crush Saga; Wordscape; Other.
    \item[\textsc{QA.8} --] \textbf{How many hours (on average) do you like to spend on mobile games?} 0--1 hour per day, 1--2 days a week; $>$1 hour per day, 1--2 days a week; 0--1 hour per day, 3--4 days a week; $>$1 hour per day, 3--4 days a week; 0--1 hour per day, every day; $>$1 hour per day, every day.
    \item[\textsc{QA.9} --] \textbf{What learning apps have you used?} LinkedIn Learning; DuoLingo; Khan Academy (spelt ``Khan Accademy'' in the form); Coursera; Headway; Brilliant; Masterclass; Athena; Udemy; Other.
    \item[\textsc{QA.10} --] \textbf{Have you used any e-learning platforms after your college/university education to learn life skills (e.g., financial literacy, time management, etc.)?} Yes; No; N/A.
    \item[\textsc{QA.11} --] \textbf{If you answered `Yes' above, which life skills have you focussed on? (Select all that apply)} Financial Literacy; Time Management; Personal Development; Communication Skills; Health and Wellness; Technical skills (e.g., coding, digital marketing); Self Awareness; Critical Thinking; Creative Thinking; Decision Making; Problem Solving; Empathy; Coping with stress; Coping with emotions; Other.
    \item[\textsc{QA.12} --] \textbf{Lifelong learning involves continuously expanding your knowledge and skills throughout your life, regardless of age or stage of career. Do you think e-learning platforms are more conducive for lifelong learning compared to traditional learning methods like classroom teaching?} Yes; No.
    \item[\textsc{QA.13} --] \textbf{Have you used online platforms for lifelong learning after graduating (e.g., upskilling)?} Yes; No; Not graduated yet.
    \item[\textsc{QA.14} --] \textbf{How long do you typically stay online during each digital learning session?} just there for the points (0--15 min); quick learning (0--15 min); while a lesson lasts (15--30 min); while a lesson lasts ($>$30 min); while i feel i am winning (30--45 min); while i feel I have upskilled (0--60 min); I like to finish my course (time doesnt matter).
    \item[\textsc{QA.15} --] \textbf{What do you like most about learning with apps? (click all that apply)} Quick and easy lessons (bite sized lessons, quick tests to check knowledge gained); Adaptive learning (changing level of questions based on my skill level); Points earned; Leaderboard; Badges; Aesthetics (music, background); Quests (challenges and awards); Personalisation (Avataars); Certification; NFT rewards (Collectables); Other.
    \item[\textsc{QA.16} --] \textbf{Gamification refers to the process of adding games or game elements to the digital learning apps, aiding you in completing your study tasks, track your lesson progress, encourage participation and assist you in competing with yourself or peers. Are you aware of gamification as a medium of engagement in learning apps?} 1 (No, what has gamification got to do with learning apps!) to 5 (Yes, I know.)
    \item[\textsc{QA.17} --] \textbf{Do you think gamification is a motivation to learn more with the apps?} 1 (Not really) to 5 (I find it an awesome tool to keep me motivated).
    \item[\textsc{QA.18} --] \textbf{To make learning more conducive with online education apps, what features do you think (or features that you might have experienced) the app would need?} Ease of use; Relevant content; Good UI/UX; Gamification; Good App support / documentation; Other.
\end{enumerate}

\section{Questionnaire: Survey B}\label{appendix:survey2}

Items and response options are reproduced from the administered Google Form (``Online Learning Behaviour -- Part 2: Have we transformed the way we learn?''). Before the questions, participants read the same information sheet as in Survey~A and gave consent in the same way. ``Other'' denotes an optional free-text field.

\begin{enumerate}[noitemsep, topsep=0pt]
    \item[\textsc{QB.1} --] \textbf{You are a} Male; Female; Prefer not to share; Other.
    \item[\textsc{QB.2} --] \textbf{School / University last attended (eg: PEC Chandigarh)} Open-ended response.
    \item[\textsc{QB.3} --] \textbf{Country of Residence (eg: Singapore)} Open-ended response.
    \item[\textsc{QB.4} --] \textbf{How young are you?} 15 to 20 years; 21 to 30 years; 31 to 40 years; Over 40 years.
    \item[\textsc{QB.5} --] \textbf{Do you use Chat GPT or any AI for any of the following?} Text generation (emails / couplets / poems / sentence correction); Image generation (describe a scene and ask GPT to generate picture); Music generation (describe a scene and ask GPT to generate a music piece); Learn new tools (eg: Tableau / excel / Flutter flow etc.); Conceptual understanding (eg: ask questions about some new concepts); Translate text (help with English or other languages like Spanish, Hindi, Japanese, etc); Coding (assistance with solving code bugs or learn new programming language); Planner (Management or Itinerary or event planning); Essay writing and evaluation; Research; Exam preparation / evaluation (students); Homework assistance or problem solving; Other.
    \item[\textsc{QB.6} --] \textbf{Which is your most favourite GPT / AI tool to use?} Chat GPT from Open AI; Gemini by Google; Co Pilot from Microsoft; Claude by Anthropic; DeepSeek; Grok by xAI (Elon Musk); Elicit by OpenAI; Other.
    \item[\textsc{QB.7} --] \textbf{What do you like most about learning with GPTs or AI? (click all that apply)} Customized Questions Answers; Immediate response; Interactive; No limit on topics; Other.
    \item[\textsc{QB.8} --] \textbf{What do you not like about learning with GPTs or AI? (click all that apply)} Response is not always correct; Limited free options; Not yet upto the mark in maths / problem solving; Adds bugs to code; Current information is not added; Difficult to use interface; Other.
\end{enumerate}

\end{appendices}

\end{document}